\documentstyle[12pt]{article}

\begin{document}

\begin{center}
\large
{\bf  Realistic lower bounds for the factorization time of large
numbers on a quantum computer}\\[.8cm]
M. B. Plenio and P. L. Knight\\[.25cm]
Optics Section, Blackett Laboratory, Imperial College,\\
London SW7 2BZ, England.\\[.25cm]
submitted to Phys. Rev. A 14. 11. 1995
\end{center}
\normalsize
\begin{abstract}
{We investigate the time $T$ a quantum computer requires to factorize
a given number dependent on the number of bits $L$ required to
represent this number. We stress the fact that in most cases one has
to take into account that the execution time of a single quantum gate
is related to the decoherence time of the qubits that are involved in
the computation. Although exhibited here only for special systems,
this inter-dependence of decoherence and computation time seems to be
a restriction in many current models for quantum computers and leads
to the result that the computation time $T$ scales much stronger with
$L$ than previously expected.}
\end{abstract}
{\bf I. Introduction}\\[.25cm]
Since Shor's discovery \cite{Shor,Ekert} of an algorithm that allows
thefactorization of a large number by a quantum computer in polynomial
time instead of an exponential time as in classical computing,
interest in the practical realization of a quantum computer has been
much enhanced. Recent advances in the preparation and manipulation of
single ions as well as the engineering of pre-selected cavity light
fields have made quantum optics that field of physics which promises
the first experimental realization of a quantum computer. Several
proposals for possible experimental implementations have been made
relying on nuclear spins, quantum dots \cite{Barenco}, cavity QED
\cite{Sleator} and on ions in linear traps \cite{Cirac}.

One can estimate the time $T$ needed for a single run of Shor's
algorithm to be equal to the time $\tau_{el}$ required to execute an
elementary logical operation multiplied by the required number of
elementary operations, which is of the form $\epsilon L^3 + O(L^2)$
\cite{Vlatko}. It should be noted that in general a single run of Shor's
algorithm will not be sufficient because it is a stochastic
algorithm. In the following we will discuss the time required to
perform one run of Shor's algorithm and if not stated explicitly
the calculation time is just the time required for this.

The calculation time has to be compared to the decoherence time
$\tau_{dec}$ of the quantum computer (eg the time in which on
average one photon will be emitted by the quantum computer). As
spontaneous emissions destroy the coherence in the quantum computer,
we need to make sure that practically no spontaneous emission occurs
during the whole computation. To ensure this, the inequality
\begin{equation}
	\tau_{dec}\gg T = \epsilon\tau_{el} L^3
	\label{1}
\end{equation}
has to be satisfied which then gives rise to an upper limit for the
numbers we are able to factorize on the quantum computer. For a
given value of $\tau_{el}$ that means that the total computation time
scales like $L^3$. To factorize a number representable by $L$ qubits,
one requires $5L+2$ qubits (in what follows we neglect the "2" here)
as work space for the necessary
calculations \cite{Vlatko}. If we assume that each qubit couples to a
different bath the decoherence time of $5L$ qubits is given by
\cite{Unruh,Palma}
\begin{equation}
	\tau_{dec} = \frac{\tau_{qb}}{5L}
	\label{1a}
\end{equation}
where $\tau_{qb}$ the decoherence time of a single qubit. The case
of qubits coupling to the same bath leads to smaller decoherence times
$\tau_{dec}$ \cite{Palma}. Combining eq. (\ref{1}) and eq. (\ref{1a})
we obtain
\begin{equation}
	\tau_{qb} \gg \tau_{el}\, 5\epsilon L^4\; .
	\label{1b}
\end{equation}
Usually $\tau_{el}$ is not assumed to be related to the decoherence
time of the quantum computer. As we will see later this is not true
in general. We will show that the dependence of the elementary time
step $\tau_{el}$ on the decoherence time $\tau_{dec}$ gives rise to
a much stronger dependence of the calculation time on the bit size
$L$. This results in a severe limitation of the maximum size of the
numbers to be factorized. In our investigation we focus on the model
put forward by Cirac and Zoller \cite{Cirac} but also show briefly
that similar restrictions apply for cavity QED implementations. We
stress that the results apply to a wide class of possible models as
most of them rely on atom-light interaction similar to that of the
models discussed here. Of course the actual form of $T(L)$ may vary
slightly from model to model.

In Section II we investigate the model of a quantum computer
proposed by Cirac and Zoller for several possible methods to store
the qubits as well as a cavity QED implementation. In Section III we
summarize our results and discuss their implications to the
realizability of quantum computers.
\\[.5cm]
{\bf II. Quantum Computation in a linear ion trap}\\[.25cm]
In the introduction we gave a simple estimate of the time $T$ a
quantum computer requires to perform Shor's algorithm. From this it
is possible to obtain an upper limit for the numbers that we are able
to factorize. However in this estimate it is usually assumed that the
execution time for an elementary logical gate does not depend on the
decoherence time of the quantum bits on which the operations are
performed. This however is not generally true. To see this note that
all the proposals for the practical implementation of quantum
computers mentioned in the introduction share a common feature. They
rely on the interaction of light with atoms where either the atoms are
used as a memory to store the qubits which are manipulated by light
fields or the light field is used as the memory which is manipulated
by the interaction with atoms. Therefore in all these schemes the
atom-light interaction represents the essential building block of
all the proposals made so far. In each of these interactions a
temporary excitation of the atoms is inevitable (even in adiabatic
excitation, given a finite excitation time) which can lead to
spontaneous decay. Obviously the interaction strength, proportional
to the Rabi frequency $\Omega$, and the spontaneous emission rate,
proportional to the Einstein coefficient of the excited level
of the transition in question, are related such that
\begin{equation}
	\Omega=\rho\Gamma^{1/2}
	\label{2}
\end{equation}
where $\Gamma$ is half the Einstein coefficient of the transition
and $\rho$ is a constant of proportionality. Certainly for a given
transition frequency $\rho$ cannot be made arbitrarily large. It is
limited due to the fact that at high intensities the two level
approximation breaks down, that the rotating wave approximation
becomes invalid and that for a sufficiently high laser intensity the
atom ionizes practically immediately. For optical transitions the
latter effect gives rise to an upper limit of the order of
\begin{equation}
	\rho_{max} \cong 10^{10} s^{1/2}\; .
	\label{3}
\end{equation}
In practize the limit will be much lower as both detuning and pulse
duration have to be controllable quantities and we have not included
the other limitations mentioned above in eq. (\ref{3}). As the
execution time $\tau_{el}$ of a quantum gate depends inversely on
the Rabi frequency $\Omega$ while the decoherence time of a qubit
$\tau_{qb}$ depends inversely on $\Gamma$ we immediately observe
via eq. (\ref{2}) that both quantities are related to each other.

In the following we will investigate how this relationship affects
the estimate for the factorization time of a number which can be
represented by $L$ qubits. First we discuss the scheme proposed by
Cirac and Zoller because it seems to be the most promising proposal.
Later we show that for cavity QED implementations similar problems
arise. In similar ways one may achieve estimates for other proposed
schemes as they mostly rely on atom-light interaction. The exact
form of $T(L)$ might be different but one will always find that the
scaling with $L$ is much stronger than expected from eq. (\ref{1}).
\begin{center}
{\bf A. Linear trap with two level atoms as qubits}
\end{center}
We now discuss the model proposed by Cirac and Zoller \cite{Cirac}.
Several ions of mass $M$ are stored in a linear trap (see Fig. 1)
and it is assumed that all translational degrees of freedom of the
ions are cooled to their respective ground state and that especially
the center-of-mass (COM) motion with frequency $\nu$ is in its
ground state. This implies that the Lamb-Dicke regime is reached.
To implement quantum gates one then applies a sequence of laser
pulses of wavelength $\lambda$ to the ions such that both the
internal degrees of freedom as well as the degree of excitation of
the COM mode may be changed. As the COM mode is a collective motion
of all ions, its excitation can be used to yield entanglement
between different ions. As an approximation it is assumed that only
the COM mode is excited because the closest lying mode has a
frequency $\sqrt{3}\nu$ and is therefore well separated from the COM
mode frequency. In the model it is assumed that the laser is detuned
such that $\Delta=-\nu$, so that the predominant contribution comes
from processes where with the excitation of the ion the COM mode is
deexcited. Processes where the ion and the mode are excited
simultaneously include rapidly oscillating phasefactors and are
neglected in the following (rotating wave approximation). One then
obtains the following Hamilton operator for an ion at the node of a
standing light field \cite{Cirac}
\begin{equation}
	H = \frac{\eta}{\sqrt{5L}}\,\frac{\Omega}{2}
	\left[ |e\rangle\langle g| a +
	|g\rangle\langle e|a^{\dagger}\right]\; .
	\label{4}
\end{equation}
where $\eta=\frac{2\pi}{\lambda}\sqrt{(\hbar/2M\nu)}\ll 1$ is the
Lamb-Dicke parameter. The $a$ and $a^{\dagger}$ are the annihilation
and creation operators of the COM mode. The Hamiltonian eq. (\ref{4})
is correct for $(\Omega/2\nu)^2\eta^2\ll 1$. This system allows the
implementation of elementary logical gates such as the controlled-NOT
gate \cite{Shor} which requires in this scheme the equivalent of four
$\pi$-pulses with the Hamiltonian eq. (\ref{4}). We use the time
required for this as a lower bound for the elementary time step
$\tau_{el}$ and find
\begin{equation}
	\tau_{el} \cong \frac{4\,\pi\sqrt{5L}}{\eta\Omega}\; .
	\label{5}
\end{equation}
Now using the fact that Shor's algorithm requires $\epsilon L^3$
elementary steps we find for the total computation time
\begin{equation}
	T \cong \frac{4\,\pi\sqrt{5L}}{\eta\Omega}\epsilon L^3\; .
	\label{6}
\end{equation}
As we want to minimize $T$, we insert the maximum value for $\Omega$
according to eq. (\ref{2}) and obtain
\begin{equation}
	T \cong \frac{4\,\pi\epsilon}{\eta\rho}
	\sqrt{\frac{5L^7}{\Gamma}}\; .
	\label{7}
\end{equation}
In this expression not all the parameters are independent, as we
have to make sure that $T$ is less than the decoherence time
$\tau_{dec}$ of the quantum computer. The decoherence time of the
quantum computer is the decoherence time of a single quantum bit
$\tau_{qb}$ divided by the number of quantum bits contained in the
quantum computer because in the course of the calculation
most of the qubits will be partially excited. We find
\begin{equation}
	\tau_{dec} = \frac{\tau_{qb}}{5L} \cong \frac{1}{5L\Gamma}
	\label{8}
\end{equation}
and obtain the inequality
\begin{equation}
	\frac{4\,\pi\epsilon}{\eta\rho}\sqrt{\frac{5L^7}{\Gamma}}
	\ll \frac{1}{5L\Gamma}\; .
	\label{9}
\end{equation}
We observe that due to eq. (\ref{2}) the decay constant of a
single qubit appears on both sides of the equation and we find
\begin{equation}
	\frac{\Gamma}{\eta^2} \ll \frac{1}{2000\pi^2}
	\left(\frac{\rho}{\epsilon}\right)^2 \frac{1}{L^9}
	\label{10}
\end{equation}
which is far more restrictive  than the estimate eq. (\ref{1b})
obtained when we assume that an elementary time step $\tau_{el}$
is independent of $\tau_{dec}$. To be able to perform Shor's
algorithm without having spontaneous emissions eq. (\ref{10}) has
to be satisfied. Using this to eliminate $\Gamma$ in eq. (\ref{7})
then gives a lower bound for the calculation time which is
\begin{equation}
	T \gg 400\pi^2 \left(\frac{\epsilon}{\rho\eta}\right)^2
	L^8\; .
	\label{11}
\end{equation}
To obtain explicit values for $T$ we assume $\eta=0.1$ and
$\rho=10^7s^{-1/2}$. The value of $\epsilon$ is of the order of
$1000$ \cite{Vlatko} so that we obtain
\begin{center}
\begin{tabular}{|c||c|c|}\hline
$L$      &      $T_{min}$        &  $\Gamma_{max}$     \\ \hline
$2$      &  $1s$            &  $10^{-1}s^{-1}$         \\ \hline
$4$      &  $259s$         &  $1.9\,10^{-4}s^{-1}$     \\ \hline
\end{tabular}\, .
\end{center}
One observes
that even with the rather large value of $\rho$ the factorization
of a $4$ bit number (eg. 15 which is the smallest composite number
for which Shor's algorithm applies \cite{Ekert}) seems to be
practically impossible when we take into account that for example
the metastable transition in Barium has a lifetime of $45s$ and
therefore $\Gamma=0.044s^{-1}$. Note that we have not taken into
account the influence of all other possible sources of error such
as counterrotating terms in the Hamilton operator, excitations of
modes other than the COM mode, errors in the pulse lengths and in the
Rabi frequencies of the pulses. One should also realize that although
a heroic experimental effort might make the factorization of a $4$
bit number possible, the factorization of any number of relevant
size seems completely out of question as the execution time of Shor's
algorithm for a $40$ bit number is $10^8$ times larger. For a $400$
bit number, which represents the upper limit which classical computers
can factorize, Shor's algorithm requires $10^{16}$ times longer than
for a $4$ bit number.

The main problem in the model seems to be that a metastable
transition cannot be driven very strongly which in turn severely
limits the execution time of an elementary gate. As a possible way
to improve the above model, it was proposed to consider a
$j=1/2\leftrightarrow j=1/2$ transition where the qubit is
represented by the two lower levels of the transition \cite{Zoller}.
However in the following we will show that this scheme suffers from
similar drawbacks as the previously investigated
system.
\begin{center}
{\bf B. The j=1/2$\leftrightarrow$ j=1/2 transition}
\end{center}
The level scheme we now investigate is depicted in Fig. 2. A qubit
is represented by the levels $1$ and $2$ which are assumed to be
stable. The transition to the two upper levels, however, may be
strong to allow for rapid transitions. As the implementation of
quantum gates requires the excitation of one phonon in the COM mode,
we need  to transfer population between the two lower levels with a
simultaneous excitation (or deexcitation) of the COM mode. To be able
to perform this population transfer without appreciable population of
the upper levels which would lead to spontaneous emissions, one has
to use the method of adiabatic population transfer \cite{adiabatic}.
The energy levels shown in Fig. 3 are the most relevant. The vertical
axis gives the energy of the bare states $|i;n\rangle$ where $i$ is
an atomic level and $n$ is the number of phonons in the COM mode.
Assume that initially the population is in level $|2;0\rangle$ and
we want to transfer it to level $|1;1\rangle$. During the
(quasi)-adiabatic population transfer one first applies a
$\sigma$-polarized laser pulse with a detuning $\Delta=-\nu$; we
assume that the ion rests at the node of the light field. The
duration of this pulse is a fixed fraction of the total length
$T_{ad}$ of the process while the length $T_{ad}$ of the process
may be varied. Later but still overlapping with the
$\sigma$-polarized laser pulse, a pulse of $\pi$-polarized light is
applied to the same ion and it is assumed that the ion is situated
at the antinode of this field. This pulse, in leading order,
preserves the excitation number of the COM mode. Again its length
is a certain fraction of the total time $T_{ad}$ and we assume that
the $\sigma$-polarized laser pulse terminates earlier than the
$\pi$-polarized pulse. If the time $T_{ad}$ in which this process
is performed is sufficiently long then the population in the upper
level $|3;0\rangle$ will be small and therefore spontaneous emissions
rare. This method certainly has the advantage that the exact pulse
shape of the laser is not as important as in the previously
discussed scheme. At first glance it also appears to be possible
that the population transfer can be made extremely fast as the Rabi
frequency is not related to the lifetime of the lower levels. However
there is a limit to the Rabi frequency. To see this we have to realize
that an adiabatic process requires infinite time. However if we want
to be able to perform the factorization in finite time we have to
take into account small deviations from the adiabatic behaviour. In
this case some population will end up in the excited levels which may
subsequently lead to spontaneous emissions. We find for the
probability $p_{em}$ that at least one spontaneous emission takes
place during the (quasi)-adiabatic process
\begin{equation}
	p_{em} \cong \beta\Gamma\frac{5L}{\eta^2\Omega_{\sigma}^2}
		     \frac{1}{T_{ad}}
	\label{13}
\end{equation}
where the constant $\beta$ depends on the peak value of the Rabi
frequency $\Omega_{\pi}$ of the $\pi$-polarized laser, the pulse
shapes and the delay between the pulses. $\Omega_{\sigma}$ is the
peak value of the Rabi frequency of the $\sigma$-polarized laser.
If $\Omega_{\pi}$ is larger than $\eta\Omega_{\sigma}$ and $\Gamma$
(which we implicitly assume in eq. (\ref{13})) we find for
$\sin^4$-pulse shapes $\beta\approx 100$. Analytically as well as
numerically one finds that $\beta$ exhibits a very slow increase
with increasing $\Omega_{\pi}$. We have assumed that the
(quasi)-adiabatic process is sufficiently slow so that the $1/T$
law applies. This is the case when the right hand side of eq.
(\ref{13}) is small compared to one. As we do not want to find
any spontaneous emission during the whole computation the
inequality
\begin{equation}
	\frac{\beta}{\eta^2}\frac{\Gamma}{\Omega_{\sigma}^2}
	\frac{5\epsilon L^4}{T_{ad}}
	= p_{em}\epsilon L^3 \ll 1
	\label{14}
\end{equation}
needs to be satisfied. This gives an estimate for the length of
an elementary time step $\tau_{el}$ which is
\begin{equation}
	\tau_{el} \approx T_{ad} \gg \frac{\beta}{\eta^2}
	\frac{\Gamma}{\Omega_{\sigma}^2}
	{5\epsilon L^4}\; .
	\label{15}
\end{equation}
Therefore we obtain for the total calculation time the estimate
\begin{equation}
	T \gg 5\beta\frac{\epsilon^2}{\eta^2}
	\frac{\Gamma}{\Omega_{\sigma}^2} L^7\; .
	\label{16}
\end{equation}
Again this estimate scales much stronger with the bitsize $L$ of
the input than expected. To see the orders of magnitude, we give
explicit values for $T$. Assuming $\eta=0.1,\beta=100,\epsilon=1000$
and $\rho=10^7s^{-1/2}$ we obtain
\begin{center}
\begin{tabular}{|c||c|}\hline
$L$      &      $T_{min}$   \\ \hline
$2$      &  $.05s$       \\ \hline
$4$      &  $6.5s$        \\ \hline
\end{tabular}
\end{center}
which indicates that even the factorization of a $4$ bit number
will be
extremely difficult to achieve, although the estimate seems to be
a little more promising than in the previous scheme. Again we have
neglected all other sources of error, such as higher order
contributions in the Lamb-Dicke parameter to the Hamilton operator
as well as counterrotating contributions neglected in the rotating
wave approximation. Because the expression eq. (\ref{16}) contains
the ratio $\Gamma/\Omega^2$, again we have similar problems as
before as this ratio cannot be made arbitrarily small.
\begin{center}
{\bf C. Cavity QED implementation}
\end{center}
Now we would like to show briefly that in cavity QED realizations
of quantum computing expressions similar to eq. (\ref{11}) and
eq. (\ref{16}) can be obtained. In cavity QED implementations of
quantum gates the atom-light interaction does not involve a
classical laser field but a quantized mode of a cavity. Before and
after the cavity we may use Ramsey zones to rotate the Bloch vector
of the atoms passing the cavity \cite{Sleator}. To perform quantum
computations such as Shor's algorithm, many cavities are required and
this obviously poses immense experimental difficulties. In the
following we neglect the restrictions arising from these problems
as well as all difficulties that arise in the realization of exactly
one atom passing with a well defined velocity through the cavity. We
will briefly show that again the lower bound for the computation
time scales much stronger than $L^3$ with the bit size $L$ of the
number to be factorized. Neglecting decay of the cavity mode, we
can estimate that the minimal computation time is of the order of
\begin{equation}
	T_{min} = \frac{\epsilon L^3}{\Omega}
	\label{17}
\end{equation}
where $\Omega$ is the Rabi frequency in the cavity-atom interaction.
While travelling in the Ramsey zones and between cavities the atoms
may decay. No decay should occur during the quantum computationcal
which leads to the condition
\begin{equation}
	\frac{\alpha\Gamma}{\Omega} \epsilon L^3 \ll 1
	\label{18}
\end{equation}
where $\alpha$ depends on the ratio between the time the ion spends
inside the cavity (where we neglect spontaneous decay) to the time
it spends outside the cavity (where it may decay). Using eq. (\ref{2})
we then obtain
\begin{equation}
	T \gg \frac{\alpha\epsilon^2 L^6}{\rho^2}\; .
	\label{19}
\end{equation}
Although this estimate seems much more promising than eq. (\ref{11})
and eq. (\ref{16}), it should be noted that it is certainly an
unrealistically low limit because we have negelcted major sources
of experimental uncertainty mentioned above. We only intend to
illustrate that again an expression similar to eq. (\ref{11}) and
eq. (\ref{16}) is found although we have discussed a completely different
realization.\\

These examples show that it seems to be a general feature that the
control of population always leads to the appearance of a factor
of the form $\Gamma/\Omega^2$ which, for a given transition
frequency, has an upper limit. There seems to be only one way
out of this dilemma. Instead of employing optical transitions to
represent qubits one could use low frequency transitions (e.g.
microwave transitions) as it was done in the cavity QED implementation
of Sleator and Weinfurter \cite{Sleator} because this can considerably
decrease the ratio $\Gamma/\Omega^2=1/\rho^2$ due to the $\omega^3$
dependence of $\Gamma$.  However as in their proposal one would need
a tremendous number of cavities it does not seem very promising. To
overcome this problem one might use the cavity field in the manner
implementation by Cirac and Zoller \cite{Cirac}. Instead of using
the COM mode to entangle different ions this task could be performed
by the cavity mode. This could be done using a linear trap to store
the ions inside a microwave cavity. This scheme then resembles that
of Sleator and Weinfurter but differs as we only require one cavity
and we do not need atomic beams with all their associated problems.
The COM mode will not be excited during the calculation as for the long
wavelength of the radiation the Lamb-Dicke parameter is extremely
small. However smaller frequencies of the incident fields mean larger
wavelengths which will make it more difficult to address single ions
with the microwave radiation. The problem of addressing a single ion,
given many are within a wavelength of the incident radiation, may be
solved by applying local magnetic or electric fields (or a suitable
field gradient) that drive all but one ion out of resonance. However
due to the small spatial separation of the ions this might be
difficult to realize experimentally. If it would be possible to
implement this idea then the lowest limit for the computation time
could become as low as eq. (\ref{19}) with a value of $\rho$ that can
be much larger than that for an optical transition. However this idea
should serve rather as a basis for discussions than a serious
proposal as we still expect the experimental difficulties to be
enormous. We are therefore not very optimistic that factorization of
nontrivial numbers will be possible in the near future.
\\[.25cm]
{\bf III. Summary}\\[.25cm]
In this paper we have investigated how the computation time which a
quantum computer needs to factorize an $L$ bit number depends on
several physical parameters. It was shown that $T$ will scale much
stronger with $L$ than previously expected. Instead of an $L^3$
dependence we find an $L^8$ or $L^7$ behaviour in the proposal of
Cirac and Zoller and $L^6$ for cavity QED realizations in which
however this limit is more of theoretical nature than of practical
importance due to other experimental problems. In the models that we
have investigated explicitly, it also turns out that the computation
time is always dependent on the ratio $\Gamma/\Omega^2$ where
$\Gamma$ and $\Omega$ are the decay constant and the Rabi frequency
of one of the transitions that are required to transfer population.
Although found for special configurations, this seems to be a general
result which limits the length of the elementary time step because
the ratio $\Gamma/\Omega^2$ cannot be made arbitrarily small for an
optical transition. As a possible way to circumvent these problems,
we briefly discussed the use of microwave transitions to store qubits
as in this case the ratio $\Gamma/\Omega^2$ becomes extremely small.
However practical problems occur which seem to make the experimental
realization of this idea difficult, although it might lead at least
to the possibility to factorize numbers which are several bits long,
a task which seems to be impossible with the present proposals.
\\[.25cm]
{\bf Acknowledgements}\\[.25cm]
We would like to thank A. Ekert and A. Barenco for discussions.
This work was supported by the Alexander-von-Humboldt Foundation,
the EC Network "Nonclassical Light", and by the UK Engineering and
Physical Sciences Research Council.

\newpage
\begin{center}
\large
{\bf FIGURE CAPTIONS}\\
\normalsize
\end{center}

\begin{description}
\begin{minipage}[t]{1.cm}\item{Fig. 1 :}\end{minipage}\hfill
\begin{minipage}[t]{14.5cm}
Schematic picture of the excitation of several ions in a linear
ion trap. The translational degrees of freedom of the ions are
assumed to be cooled to their respective ground states. To
implement quantum gates, standing wave fields interact with the
ions and thereby changing the inner state of the ions as well as
the state of the center-of-mass mode which leads to entanglement.\\
\end{minipage}
\\
\begin{minipage}[t]{1.cm}\item{Fig. 2 :}\end{minipage}\hfill
\begin{minipage}[t]{14.5cm}
A $j=1/2\leftrightarrow j=1/2$ transition. The qubit is
represented by the two lower levels $1$ and $2$. Population transfer
requires two different lasers. Adiabatic population transfer
minimizes unwanted population in the upper level.\\
\end{minipage}
\\
\begin{minipage}[t]{1.cm}\item{Fig. 3 :}\end{minipage}\hfill
\begin{minipage}[t]{14.5cm}
The $j=1/2\leftrightarrow j=1/2$ transition including the quantized
center-of-mass motion. $|i;n\rangle$ denotes an atomic level $i$
and $n$ phonon in the center-of-mass mode. For the implementation of
a controlled-NOT gate we need to be able to transfer population from
state $|2;0\rangle$ to state $|1;1\rangle$ and vice versa. To
minimize population in the excited levels population transfer is
performed using adiabatic population transfer with counterintuitive
pulse sequence.\\
\end{minipage}
\end{description}

\newpage
\begin{figure}[hbt]
\newcounter{cms}
\setlength{\unitlength}{1.0mm}
\begin{picture}(100,50)
\put(0,55){\makebox(5,5)[bl]{Fig. 1\hspace*{5cm} Plenio
\hspace*{1.cm}PRA}}
\thicklines
\put(0,0){\line(1,0){10}}
\put(0,0){\line(0,1){5}}
\put(0,5){\line(1,0){10}}
\put(10,2.5){\oval(10,5)[r]}
\multiput(25,2.5)(15,0){8}{\circle*{5}}
\multiput(25,-3)(15,0){8}{\vector(1,0){5}}
\multiput(25,-3)(15,0){8}{\vector(-1,0){5}}
\multiput(23,20)(15,0){8}{\line(0,1){10}}
\multiput(27,20)(15,0){8}{\line(0,1){10}}
\multiput(23,30)(15,0){8}{\line(1,0){4}}
\multiput(23,20)(15,0){8}{\line(1,0){1}}
\multiput(26,20)(15,0){8}{\line(1,0){1}}
\multiput(39.5,29)(.2,0){5}{\line(0,-1){22}}
\multiput(99.5,29)(.2,0){5}{\line(0,-1){22}}
\large
\put(49,35){\makebox(5,5)[bl]{Standing wave laser fields}}
\normalsize
\put(145,2.5){\oval(10,5)[l]}
\put(145,0){\line(1,0){10}}
\put(155,0){\line(0,1){5}}
\put(155,5){\line(-1,0){10}}
\end{picture}\\[1.5cm]
\end{figure}
\newpage
\begin{figure}[hbt]
\setlength{\unitlength}{.9mm}
\begin{picture}(80,60)
\put(0,60){\makebox(5,5)[bl]{Fig. 2\hspace*{5cm} Plenio
\hspace*{1.cm}PRA}}
\thicklines
\put(50,0){\line(1,0){30}}
\put(62.5,-8){\makebox(5,5)[bl]{$|1\rangle$}}
\put(50,40){\line(1,0){30}}
\put(62.5,45){\makebox(5,5)[bl]{$|3\rangle$}}
\put(90,0){\line(1,0){30}}
\put(102.5,-8){\makebox(5,5)[bl]{$|2\rangle$}}
\put(90,40){\line(1,0){30}}
\put(102.5,45){\makebox(5,5)[bl]{$|4\rangle$}}
\put(105,0){\circle*{4}}
\put(105,38){\vector(0,-1){34}}
\put(105,4){\vector(0,1){34}}
\put(65,3){\vector(1,1){35}}
\put(99,37){\vector(-1,-1){35}}
\thinlines
\large
\put(107,16){\makebox(5,5)[bl]{$\pi$}}
\put(73,16){\makebox(5,5)[bl]{$\sigma$}}
\end{picture}\\[2.cm]
\end{figure}
\newpage
\begin{figure}[hbt]
\setlength{\unitlength}{1.0mm}
\thicklines
\begin{picture}(50,60)
\put(0,65){\makebox(5,5)[bl]{Fig. 3\hspace*{5cm} Plenio
\hspace*{1.cm}PRA}}
\put(0,0){\line(1,0){30}}
\put(11,-6){\makebox(5,5)[bl]{$|1;0\rangle$}}
\put(0,40){\line(1,0){30}}
\put(11,45){\makebox(5,5)[bl]{$|3;0\rangle$}}
\put(40,0){\line(1,0){30}}
\put(51,-6){\makebox(5,5)[bl]{$|2;0\rangle$}}
\put(40,40){\line(1,0){30}}
\put(51,45){\makebox(5,5)[bl]{$|4;0\rangle$}}
\put(85,10){\line(1,0){30}}
\put(96,5){\makebox(5,5)[bl]{$|1;1\rangle$}}
\put(85,50){\line(1,0){30}}
\put(96,55){\makebox(5,5)[bl]{$|3;1\rangle$}}
\put(125,10){\line(1,0){30}}
\put(136,5){\makebox(5,5)[bl]{$|2;1\rangle$}}
\put(125,50){\line(1,0){30}}
\put(136,55){\makebox(5,5)[bl]{$|4;1\rangle$}}
\thicklines
\put(55,0){\circle*{4}}
\put(60,37){\vector(3,-2){38}}
\put(97,12.33){\vector(-3,2){38}}
\put(55,38){\vector(0,-1){34}}
\put(55,4){\vector(0,1){34}}
\thinlines
\large
\put(57,20){\makebox(5,5)[bl]{$\pi$}}
\put(81,25){\makebox(5,5)[bl]{$\sigma$}}
\end{picture}
\end{figure}


\begin{thebibliography}{99}
\bibitem{Shor} P. W. Shor, in {\em Proceedings of the 35th Annual
Symposium on the Foundations of Computer Science, Los Alamitos,}
CA (IEEE Computer Society Press, New York, 1994), p. 124
\bibitem{Ekert} A. Ekert and R. Josza, {\em Shor's Quantum
Algorithm for Factorising Numbers}, preprint to appear in Rev.
Mod. Phys.
\bibitem{Barenco} A. Barenco, D. Deutsch, A. Ekert, and R. Josza,
Phys. Rev. Lett. {\bf 74}, 4083 (1995)
\bibitem{Sleator} T. Sleator and H. Weinfurter, Phys. Rev. Lett.
{\bf 74}, 4087 (1995)
\bibitem{Cirac} J. I. Cirac and P. Zoller, Phys. Rev. Lett.
{\bf 74}, 4091 (1995)
\bibitem{Vlatko} V. Vedral, A. Barenco, and A. Ekert, {\em Quantum
Networks for Elementary Arithmetic Operations} submitted to
Phys. Rev. A
\bibitem{Unruh} W. G. Unruh, Phys. Rev. A {\bf 51}, 992 (1995)
\bibitem{Palma} G. M. Palma, K.-A. Suominen, and A. Ekert, {\em
Quantum Computers and Dissipation} submitted to Proc. Roy. Soc.
London 1995
\bibitem{Zoller} P. Zoller, {\em Lectures given at Les Houches
Summer School on Quantum Fluctuations 1995}, (Elsevier Publishers,
Amsterdam, in press 1995) eds. S. Reynaud et al.
\bibitem{adiabatic} B. W. Shore, J. Martin, M. P. Fewell, and
K. Bergmann, Phys. Rev. A {\bf 52}, 566 (1995)\\
J. Martin, B. W. Shore, and K. Bergmann, Phys. Rev. A {\bf 52},
583 (1995) and references therein
\end{thebibliography}
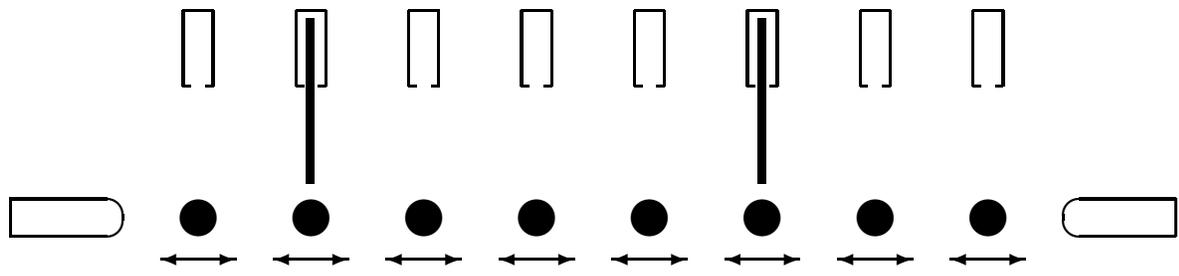
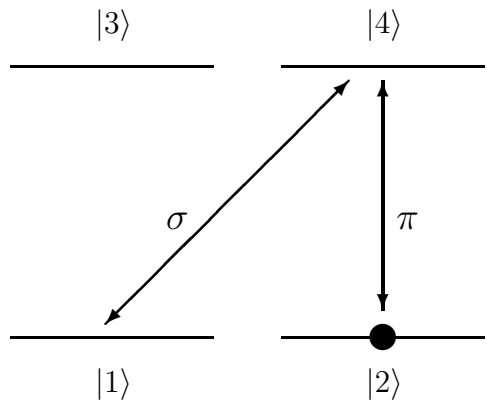
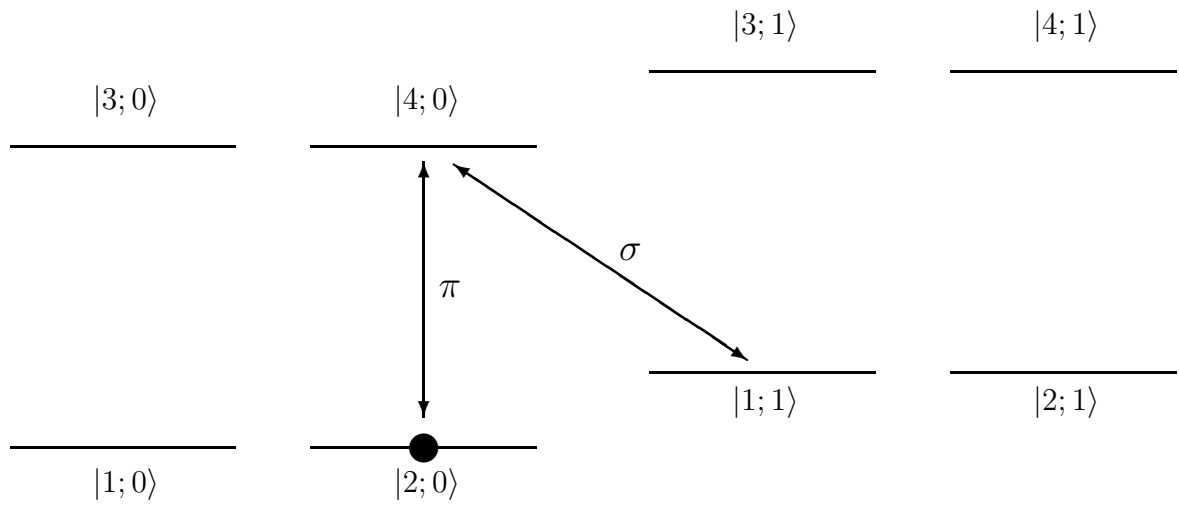
\end{document}